# Investigation of the atomic coordinates of $CeNiC_2$ under pressure: switching of the Ce-Ce first nearest neighbor direction

Hanming Ma[1], Dilip Bhoi[1], Jun Gouchi[1], Hiroyasu Sato[2], Toru Shigeoka[3], J.-G. Cheng[4], and Yoshiya Uwatoko[1]

[1]*Instittute for Solid State Physics, The University of Tokyo, Kashiwanoha 5-1-5, Kashiwa, Chiba 277-8581, Japan*

[2]*Rigaku Corporation, Akishima, 196-8666, Japan*

[3]*Graduate School of Science Technology, Yamaguchi University, Yamaguchi, 753-8512, Japan*

[4]*Beijing National Laboratory for Condensed Matter Physics and Institute of Physics, Chinese Academy of Sciences, Beijing 100190, China*

When pressurized, the heavy fermion compound $CeNiC_2$ reveals a rich electronic phase diagram and shows unconventional superconductivity with a transition temperature $T_c \sim 3.7$ K, the highest among Ce-based heavy fermion superconductors [S. Katano et al., Phys. Rev. B. 99, 100501(R) (2019)]. Understanding of this appearance of superconductivity in the vicinity of magnetic quantum critical point is still lacking. Given that physical properties of $CeNiC_2$ are sensitive to subtle changes in the interatomic distances, information on atomic coordinates may offer essential insights into the local lattice arrangements, thus the mechanisms behind the exotic phases and phase transitions. However, extraction of precise information on the atomic coordinates under pressure remains a challenge. To find a correlation between the local lattice environments and exotic physical properties in $CeNiC_2$, we investigate its crystal structure from ambient pressure to 18.6 GPa via single crystal X-ray diffraction. The pressure dependence of lattice parameters reveals anisotropic linear compressibility, $k$, following the relationship "$|k_a|$ ($3.70\times10^{-3}$ GPa$^{-1}$) > $|k_c|$ ($1.97\times10^{-3}$ GPa$^{-1}$) > $|k_b|$ ($1.39\times10^{-3}$ GPa$^{-1}$)", and a large bulk modulus, $B_0 \sim 134$ GPa. Although the atomic coordinates between Ce and Ni remain unchanged under applied pressure, direction of the first nearest and the second nearest neighbors between both the Ce-Ce and Ni-Ni atoms switch ~ 7 GPa. Notably, this is the same pressure that antiferromagnetic ordering temperature reaches maximum in the pressure temperature phase diagram of $CeNiC_2$. Our results suggest that the direction of nearest neighbors interchange might play a key role in the suppression of magnetic order and the enhancement of Kondo effect.

## Introduction

Rare earth compounds host a wealth of exotic quantum properties due to the interaction between magnetic (*f*) and conduction (*d* and *s*) electrons, such as heavy fermion state, quantum criticality, and unconventional superconductivity [1]. In these materials, the competing interplay between the long-range RKKY interaction and the Kondo effect can be susceptible to subtle changes in the interatomic distance. Thus, microscopic structural information, such as atomic coordinates, under external tuning may provide critical insights into the mechanisms behind the novel phases and phase transitions comprising their phase diagrams [2,3,4,5]. A typical example is the $R$NiC$_2$ family (where $R$ = elements in lanthanide series), for which decreasing atomic radii of $R$ elements leads to several competing states with different energy scales, including superconductivity (SC), charge density wave (CDW) and magnetic ordering of $R$ elements [6]. The $R$NiC$_2$ compounds crystallize in a non-centrosymmetric (NCS) orthorhombic structure with space group *Amm2*. It consists of alternatively stacked $R$ and NiC$_2$ atomic layers along the $a$-axis and breaks the inversion symmetry along the $c$-axis, as shown in Fig. 1(a) for CeNiC$_2$ as a representative. Among these materials, only LaNiC$_2$ displays SC below $T_c$ ~ 2.7 K at ambient conditions [6]. Further, upon pressurizing, $T_c$ reveals a dome-like shape with a maximum ~ 3.5 K around 3.5 GPa [7]. Close to this pressure, a different state with high energy scale emerges and competes with SC state. However, other $R$NiC$_2$ compounds, with the $R$ element radii smaller than Ce, display CDW order and the CDW transition temperature increasing with the decreasing atomic size of the $R$ element [6].

Among the $R$NiC$_2$ compounds, CeNiC$_2$ features unique properties such as heavy fermionic behavior and multiple magnetic ordering. With decreasing temperature, CeNiC$_2$ displays an incommensurate antiferromagnetic transition (ICAF) at $T_{\mathrm{ICAF}}$ ~ 20 K followed by commensurate antiferromagnetic transition at $T_{\mathrm{CAF}}$ ~ 10 K and a ferromagnetic ordering below 2.2 K [8]. The $T_{\mathrm{ICAF}}$ increases with increasing applied pressure, reaching a maximum around 7 GPa [8]. As pressure is increased further to 11 GPa, the ICAF order is entirely suppressed, and a SC dome with a maximum $T_c$ ~ 3.5 K emerges in a narrow pressure range around 11 GPa [8]. Furthermore, this SC state exhibits a large upper critical field, $H_{c2}(0)$ ~18 T, almost three times higher than the Pauli paramagnetic limiting field, $H_p$ ~ 6.5 T, suggesting unconventional nature of the SC state [8]. It is worth noting that, $T_c$ of CeNiC$_2$ is the highest among all known Ce-based heavy fermion superconductors [8].

In the NCS superconductors, antisymmetric spin-orbit coupling may occur due to the lack of inversion symmetry. Under certain conditions, spin-triplet cooper pairing with large $H_{c2}(0)$ is favored, like in the case of CePt$_3$Si [9]. Moreover, NCS superconductors are expected to host a mixing of spin triplet and singlet pairing [10]. Although CeNiC$_2$ possess NCS crystal structure at ambient pressure, it is not clear whether the NCS structure is maintained across the pressure range where SC state appears. Apart from the symmetry of the underlying crystal structure, the subtle variation of interatomic distances under pressure can play a crucial role in shaping the macroscopic physical properties of a heavy fermion material. For example, in Ce$T_2X_2$ ($T$ = transition metals, $X$ = $p$-like metal), it was shown that the interatomic distances sensitively affect the magnetic order and the strength of Kondo coupling between conduction electrons and the $f$ electrons [11,12]. Therefore, structural information under pressure is highly desirable for understanding the magnetic and electronic properties of $CeNiC_2$. However, extraction of the atomic coordinates with high precision under pressure is a difficult task.

In this work, therefore, we investigate the effect of pressure on atomic coordinates of $CeNiC_2$, employing single crystal x-ray diffraction (XRD) measurements at room temperature up to 18.6 GPa. The pressure dependences of *a*-, *b*-, and *c*-lattice reveal anisotropic linear compressibility, *k,* and a large bulk modulus, $B_0 \sim 134$ GPa. The *k* is highest along the *a*-axis, and lowest along the *b*-axis, whereas along the *c*-axis it is in between the *a*- and *b*-axis. Moreover, our data also reveals that the interchange of the first and the second nearest direction of both Ce-Ce and Ni-Ni atoms occurs ~ 7 GPa. Also, we observed the nonmonotonic pressure dependence of interatomic distance between C-Ce, C-Ni, and C-C atoms at ~ 7 GPa.

## Experimental methods and techniques

High-quality single crystals of $CeNiC_2$ were grown by Czochralski pulling method using a tetra-arc furnace in Argon gas environments. For the crystal synthesis, we used Ce, Ni, and C with purities 4N (Ames Lab.), 4N8 (Koujundo Chemical Laboratory Co., Ltd.), and 5N5 (Shoyoudenko Co., Ltd.), respectively. Single crystal X-ray diffraction at 293(2) K was performed using a Rigaku XtaLab MicroMax007 HFMR (Mo-K$\alpha$ radiation ($\lambda$ = 0.71073 Å)) with a HyPix-6000 diffractometer. Using Olex2 [13], the structure was solved by direct methods (SHELXT 2018/2) [14] and refined by full-matrix least-squares methods on $F^2$ values (SHELXL 2018/3) [14]. High pressure experiments were performed using a diamond anvil cell (DAC) equipped with 300 μm culet size. A Rhenium metal gasket with an initial thickness of 200 μm was pre-indented to ~ 60 μm. Subsequently, a 160 μm hole inside the gasket is drilled. A piece of carefully selected $CeNiC_2$ single crystal with a typical size of ~100 μm was loaded in DAC together with a piece of ruby as a pressure manometer. A mixture of methanol and ethanol alcohol (4:1) was used as a pressure transmitting medium. The pressure inside the DAC was determined by the ruby fluorescence calibration method [15].

## Results

Fig. 1(b) shows the diffraction pattern of a $CeNiC_2$ single crystal in the (*h*, *k*, 0) plane at ambient condition. The sharp diffraction pattern indicates the high crystallinity of the samples. The pattern was successfully refined using an orthorhombic structure in *Amm2* space group with $R_1$ = 1.67 %. The estimated lattice parameters $a$ = 3.8742(2) Å, $b$ = 4.5459(2) Å, $c$ = 6.1590(3) Å, and cell volume $V$ = 108.471(11) Å$^3$, are in good agreement with the previous report [16]. In Fig. 1(c) and Fig. 1(d), we show the diffraction patterns of the $CeNiC_2$ crystal at 0 GPa inside the DAC and at 18.6 GPa as a representative pattern for high pressure. As the DAC put a constraint on the open angle, the number of accessible diffraction peaks are limited. Despite this and the large background of DAC, clear and sharp diffraction spots remain visible without any expansion in Fig. 1(c) and Fig. 1(d). This indicates the crystal is not crushed by pressure and good crystallinity is maintained up to 18.6 GPa.

The lattice parameters and cell volume at 0 GPa for the sample inside the DAC are estimated as $a$ = 3.8753(13) Å, $b$ = 4.5394(12) Å, $c$ = 6.1609(15) Å, and $V$ = 108.380(5) Å$^3$, respectively. Although these deduced values are slightly different due to the different sample setting conditions inside the DAC, the differences are negligible, indicating the reliability of our measurements. The pressure dependent lattice parameters and unit cell volumes with varying pressure are plotted in Fig. 1(e) revealing a linear decrease with pressure. The pressure dependence of lattice parameters and cell volume in $CeNiC_2$ under pressure are summarized in Supplementary TABLE I. These results suggest that $CeNiC_2$ does not show any signs of crystal structural phase transition across the measured pressure range.

Fig. 1(f) shows the normalized lattice parameters and unit cell volume of $CeNiC_2$ in the pressure range from 0 to 18.6 GPa. The normalized lattice parameters show anisotropic behaviors, the $a$-axis decreases at a much faster rate compared to $b$- and $c$-axis. Using a linear fit to the pressure dependence of lattice parameters, the compressibility along each axis is calculated as $k_a$ = d($a/a_0$)/d$P$ = -3.70×10$^{-3}$ GPa$^{-1}$, $k_b$ = d($b/b_0$)/d$P$ = -1.39×10$^{-3}$ GPa$^{-1}$, and $k_c$ = d($c/c_0$)/d$P$ = -1.97×10$^{-3}$ GPa$^{-1}$, respectively. The lattice compressibility along $a$-axis is the highest, whereas $b$-axis is the lowest.

To estimate the Bulk modulus, $B_0$, of the $CeNiC_2$, we fit the normalized cell volume as shown in Fig. 1 (h) by the Birch-Murnaghan equation of state (BM-EoS) [17], expressed as:

$$\frac{V(P)}{V_0} = \left(\left(\frac{B_0'}{B_0}\right)P + 1\right)^{-\frac{1}{B_0'}} \qquad (1),$$

where $V$ is the volume at a fixed pressure $P$, $V_0$ is the volume at ambient pressure, and $B_0$' is the first derivative of the bulk modulus. From the fitting of BM-EoS equation in the pressure range 0 to 18.6 GPa, we obtain $B_0$ ~ 134 GPa with $B_0$' = 0.75. In the Ce-based heavy fermion superconductors the bulk modulus typically varies in the range of 60 to 140 GPa [18,19,20,21,22,23,24]. Our results suggest that the $B_0$ of $CeNiC_2$ is one of the highest among the Ce-based heavy fermion compounds. Compared to other Ce based compounds, the $B_0$' value in $CeNiC_2$ is quite low [25]. This indicates the bulk modulus of $CeNiC_2$ under pressure is almost constant within the measurement pressure range.

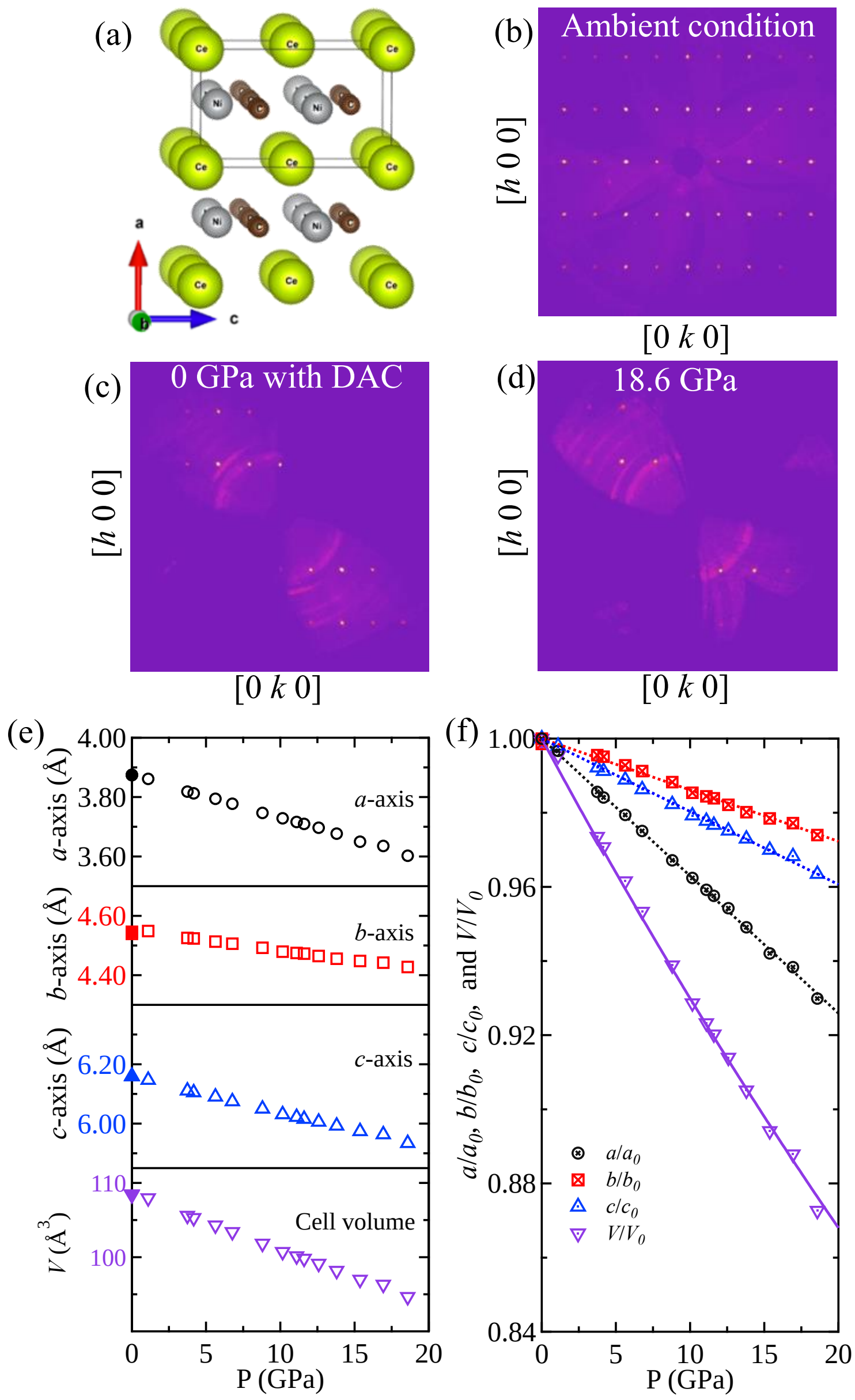

**FIG. 1:** (a) The crystal structure of $CeNiC_2$ showing the stacking of the Ce and $NiC_2$ layers along the *a*-axis. The rectangular box illustrates the unit cell. The diffraction pattern of a $CeNiC_2$ single crystal in (*h*, *k*, 0) plane at (b) ambient condition and (c) for sample inside the DAC at 0 GPa. (d) The diffraction pattern at 18.6 GPa. (e) The pressure dependence of lattice parameters and unit cell volume. The solid symbols represent the data obtained at ambient conditions. (f) The pressure dependence of normalized lattice parameters and unit cell volume. The error bars are smaller than the symbols. The dashed lines are the linear fitting to the pressure dependence of lattice parameters used for estimating the compressibility. The solid line represents a fit of the Birch-Murnaghan equation of state to the normalized unit cell volume.

To obtain further insight, the pressure dependent atomic coordinates of Ce (0, 0, *Z*), Ni (0.5, 0, *Z*), C (0.5, *Y*, *Z*), are plotted in Fig. 2(a). For clarity, Fig. 2(b) illustrates the positions of different atoms in the unit cell, circled by dashed lines. The Ce atom occupies the corner sharing and the face center in (0 1 1) plane. Ni atom and C-C dimer are located in the *a*/2 plane with the alternate stacking of Ce and $NiC_2$ layers along the *a*-axis. The pressure dependence of atomic coordinates and equivalent isotropic displacement parameters in $CeNiC_2$ under pressure are summarized in Supplementary TABLE II. The difference between the ambient condition and that obtained from the measurement inside the DAC at 0 GPa are almost less than 0.0006 in *Z* factor of Ce, Ni, and C respectively. On the other hand, a relatively large difference appears in *Y* factor of C, about 0.026. This difference possibly arises as the energy of the laboratory X-ray source used in this work is not as high energy as neutron or synchrotron source. Further, due to the limitation imposed by the DAC, the accessible diffraction points are limited. In this situation, the atomic coordinate of light element, like carbon, is difficult to obtain as reliably as Ni or Ce. Nevertheless, considering a large error bar (~ 10%), we still detect a nonmonotonic pressure dependence of the *Y* factor of C atom as in Fig. 2(a). From 0 to 7 GPa, the *Y* factor decreases as pressure increase. Whereas over 7 GPa, it remains almost pressure independent. Compared to *Y* factor of C atom, the *Z* factor of C as well as Ni and Ce atoms varies negligibly with pressurizing. This indicates that the bond length of the C-C dimers varies obviously under pressure.

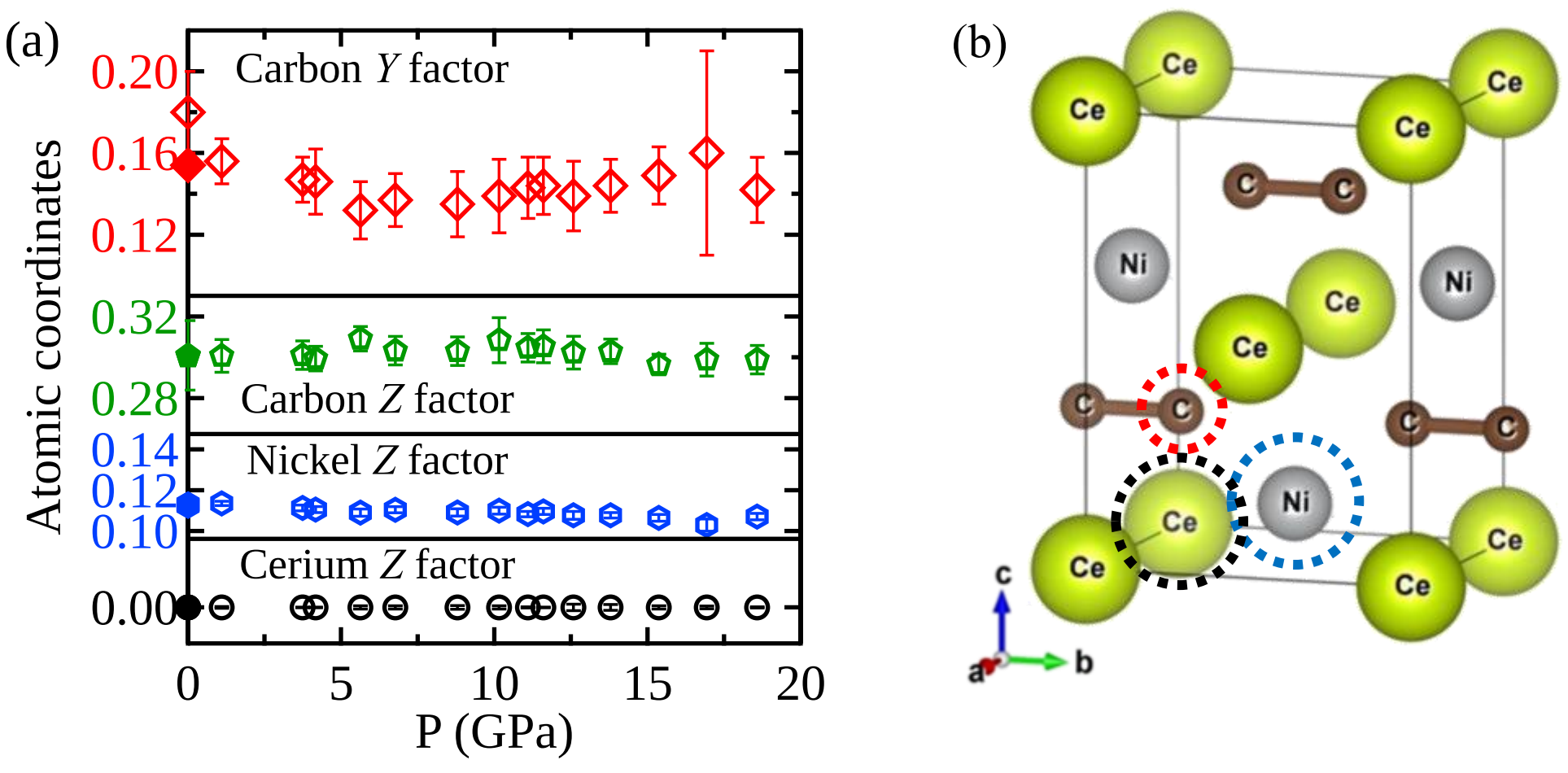

**FIG. 2:** (a) The *Y* atomic factor of C atom and the *Z* factor of C, Ni and Ce atom under pressure. The solid symbol represents the data at ambient condition. (b) The positions of C, Ni, and Ce atoms in the unit cell are illustrated with red, blue, and black dashed circles respectively.

The detailed knowledge of the atomic coordinates allowed us to estimate the pressure dependence of different interatomic distances. Fig. 3(a) and Fig. 3(b), show the interatomic distances between Ce-Ce and Ni-Ni atoms. At ambient conditions, the interatomic distance 1 is longer than 2, for both Ce-Ce and Ni-Ni atoms. Thus, before 7 GPa, the interatomic distance 1 is the direction of the first nearest neighbor (FNN), which is in the *bc* plane along [0,1,1] direction. And the interatomic distance 2 is the second nearest neighbor (SNN) direction, which is along the *a*-axis. Interestingly, the interatomic distances 1 and 2 become comparable near 7 GPa. While for pressure higher than 7 GPa, the interatomic distances 1 become shorter than 2. This suggests that over 7 GPa, the FNN direction lie along the *a*-axis, whereas the SNN direction lie along [0,1,1]. This is interesting because the atomic coordinates of Ce and Ni are almost pressure independent. But due to the anisotropic compressibility of the lattice, the FNN and SNN direction interchanges near 7 GPa.

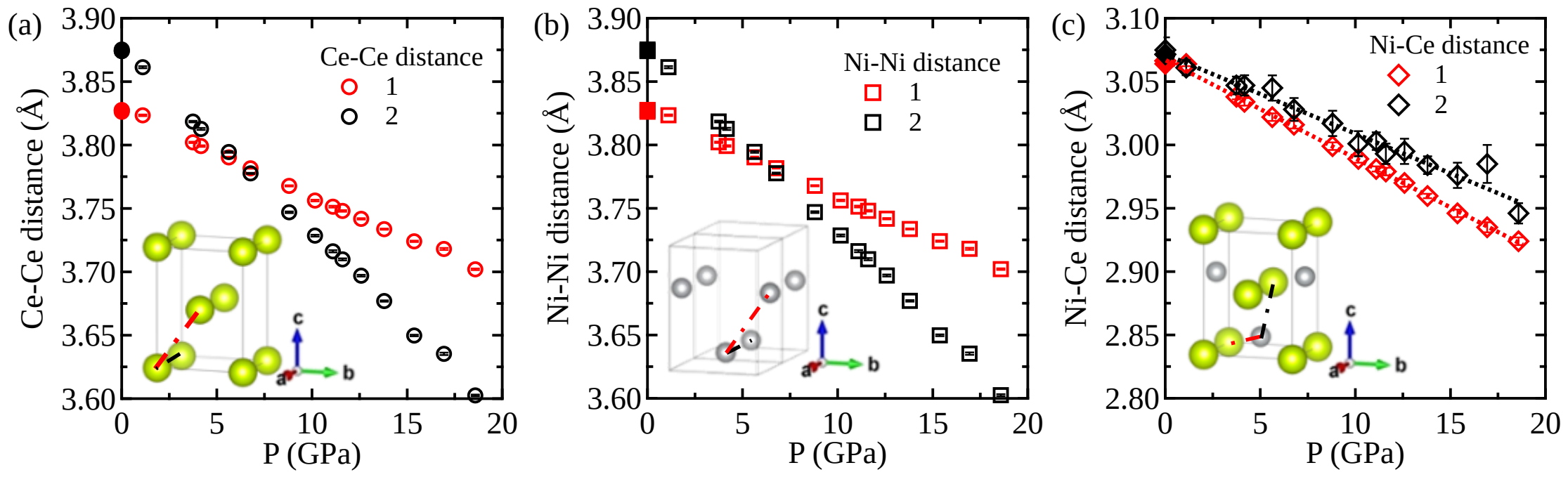

**FIG. 3:** The pressure dependence of interatomic distances between (a) Ce-Ce atom, (b) Ni-Ni atom, and (c) Ni-Ce atom. The dash lines in (c) are the linear fitting results. The red and black dot dashed lines in the insert illustrate the direction of interatomic distance 1 and 2 in the unit cell. The solid symbol shows the ambient condition data.

Fig. 3(c) shows the pressure dependence of interatomic distance between the Ni-Ce atoms. As shown in the figure, the interatomic distances 1 (FNN) and 2 (SNN) are almost identical at ambient conditions, and pointing along the [0.5, 0.5, 0.11298] and [0.5, 0, 0.38712] directions, respectively. The interatomic distances 1 and 2 follow a linear decrease from 0 to 18.6 GPa. Interestingly, due to the anisotropic compressibility, the interatomic distance difference between FNN and SNN becomes higher with pressurizing.

Fig. 4(a) shows the pressure dependence of C-C bond length, it reveals a nonmonotonic pressure dependence similar to the *Y* factor of the C atom. The estimated C-C bond length at 0 GPa, for the sample inside the DAC, is 0.2 Å larger than the ambient condition but falls within the error bar. Whereas, even taking into consideration of the large error bar, a linear fitting in the whole pressure range could not describe the data. From 0 to 7 GPa, the C-C bond length decreases with pressurizing, while pressure higher than 7 GPa the C-C bond length slightly increase, suggesting that C-C bond length passes through a minimum near 7 GPa.

Similarly, we also analyzed the pressure dependence of the interatomic distances between C-Ni and C-Ce atoms, as in Fig. 4(b) and Fig. 4(c) respectively. There are two comparable interatomic distances between C-Ni atoms and C-Ce atoms as shown in Fig. 4(d). For C-Ni, the interatomic distance 1 increases with pressure from 0 to 7 GPa, and it slightly decreases while further pressurizing. Whereas interatomic distance 2 decreases from 0 to 7 GPa, and it is almost pressure independent under higher pressure. For C-Ce, the interatomic distance 1 is almost pressure independent, but for pressure higher than 7 GPa it decreases with pressurizing, showing a clear anomaly around 7 GPa. On the other hand, interatomic distance 2 decreases linearly with pressure from 0 to 18.6 GPa. Above 7 GPa, the C-Ce interatomic distances 1 and 2 become almost comparable.

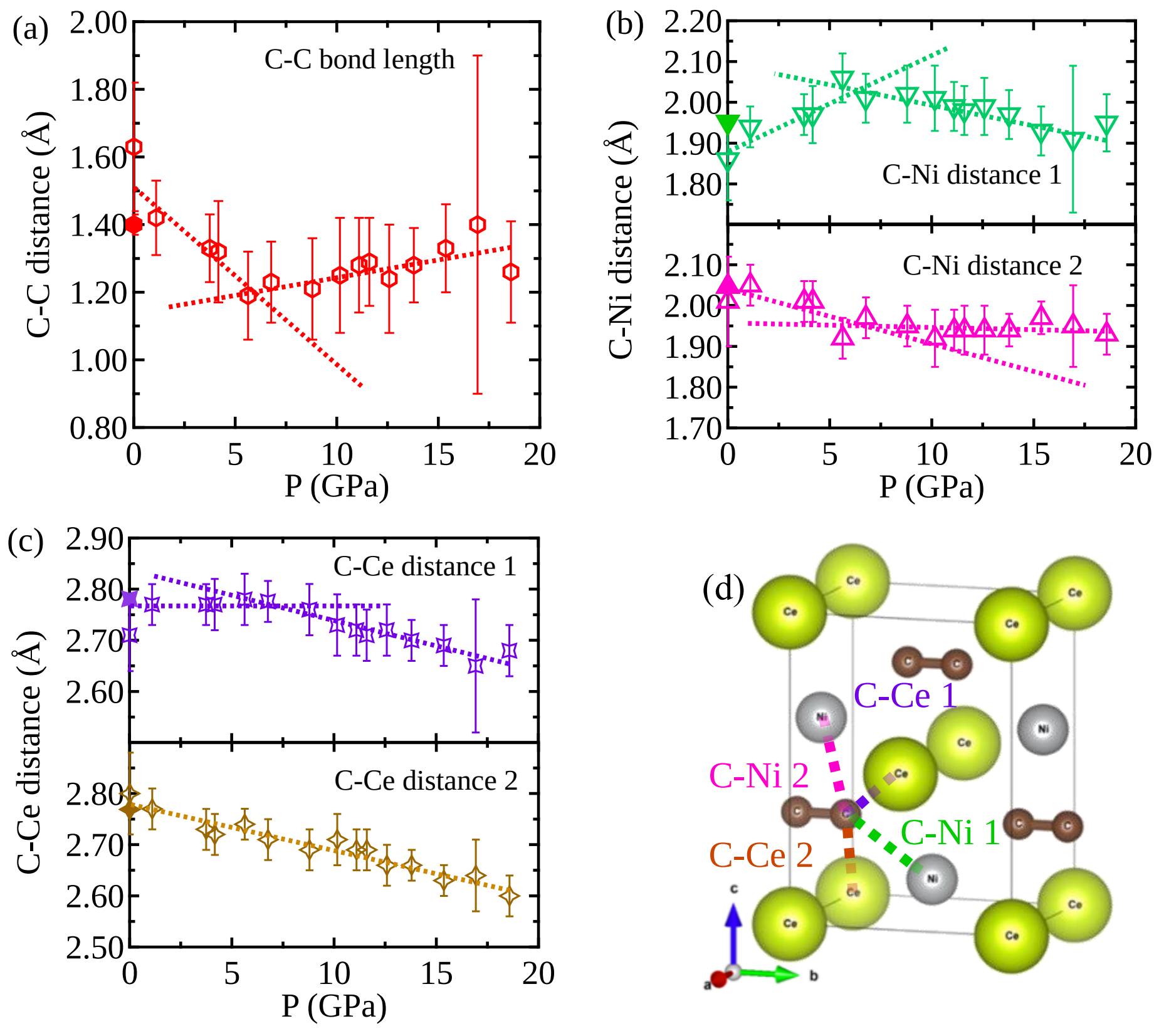

**FIG. 4:** The pressure dependence of (a) C-C bond length. Interatomic distance between the (b) C-Ni atom, and (c) C-Ce atom under different pressure. The solid symbols represent the data at ambient condition. (d) The illustration of different interatomic distances in the $CeNiC_2$ unit cell.

## Discussions

For Ce based heavy fermion compounds, it is well known that the separation distance *r* between the Ce atoms plays a sensitive role in deciding the nature of the ground state, as both the RKKY interaction and the Kondo effect are primarily governed by the interaction between *f*- and conduction electron. In the Anderson model, the energy scales of RKKY interaction ($k_B T_{RKKY}$) and Kondo effect ($k_B T_K$) are related to the strength of the *c-f* interaction ($J_{cf}$) [26], as expressed below:

$$k_B\, T_{RKKY} \propto J_{cf}^2\, D(E_F)\, F(2k_F r) \qquad (2),$$

$$k_B\, T_K \propto W \exp\,(-1/\, J_{cf}\, D(E_F)) \qquad (3),$$

$$J_{cf} \propto |V_{cf}|^2\, /\, (E_F - E_{4f}) \qquad (4),$$

where $D(E_F)$ is the density of states of conduction electrons at the Fermi level $E_F$, $F(2k_F r)$ is Friedel oscillation function of distance $r$ between two localized spins, and $k_F$ Fermi momentum, $W$ is the conduction electron band width, $V_{cf}$ is a matrix element of 4*f* electrons and conduction electron coupling, and $E_{4f}$ is the 4*f* electron energy level. The nature of the RKKY interaction follows the Friedel oscillation function $F(2k_F r)$, which is sensitive to the $r$ and $k_F$ [26]. The value of $F(2k_F r)$ determines the type of magnetic order: a positive value leads to an FM order, while a negative value results in an AFM order. According to the neutron diffraction [27] $CeNiC_2$ exhibits a complex 3D magnetic ordering at ambient pressure with the moments pointing along the *b*-axis. The FNN Ce-Ce atoms lie in the *bc* plane, along [0,1,1], suggesting the dominant magnetic exchange interaction is in the Ce atom layer. While the SNN lies along the *a*-axis, which is between the Ce atom in adjacent layers. At 0 GPa, the difference between the FNN and SNN distance is about ~ 0.05 Å. As pressure increases, the FNN and SNN interatomic distance of Ce-Ce atoms decrease and become equal around 7 GPa. However, above 7 GPa, the direction of FNN and SNN interchange; FNN lies along the *a*-axis, between the Ce atom in adjacent layers, while SNN lies along the *bc* plane, in the Ce atom layer. Moreover, with further increase in pressure the difference in the FNN and SNN increases and becomes twice at 18.6 GPa. Coincidentally, the ICAF order starts to weaken above 7 GPa with the appearance of Kondo effect which increases with increasing pressure [8], suggesting that the spin structure of $CeNiC_2$ is strongly influenced by the increase in the interplane interaction between the Ce atoms in different layers. It is likely that with pressure increase, as the FNN and SNN interatomic distance started to shrink, both the in-plane and inter-plane the magnetic exchange interaction between the Ce atoms strengthens. Whereas the decrease of ICAF order above 7 GPa [8] clearly suggests that the Ce-Ce exchange interaction in the *bc* plane and along the *a*-axis become comparable and compete. It should be noted that the interatomic distances sensitively affect the magnetic structure in $CeT_2X_2$ compounds [11, 12]. In $CeRh_2Ge_2$, it was observed that both the Ce-Ce in-plane and inter-plane interatomic distance have significant effect on magnetic exchange interaction, thus strongly influencing the spin structure [28]. Similarly, in $CeCu_2Ge_2$, the Ce-Cu distances was thought govern the $J_{cf}$ under pressure [29].

This anomalous direction interchange of FNN and SNN possibly related to the anisotropic compressibility of $CeNiC_2$. The compressibility along *b*- and *c*-axis are comparable and much lower than that along the *a*-axis, suggesting that the stiffness of $NiC_2$ layer hinders the compression along the *b*- and *c*-axis. In a previous work [30], the stiffness of the $NiC_2$ layer was attributed to the strong C-C bond and Ni-C interaction in the *bc* plane. This indicates that the interaction within $NiC_2$ layer is stronger than the interaction between Ce and $NiC_2$ layers, resulting in anisotropic compressibility. Besides, the C-C, C-Ni, and C-Ce interatomic distances shows nonmonotonic pressure dependence and reveal anomalies around 7 GPa. Such nonmonotonic pressure dependencies may be the consequence of increasing Ce-Ce interaction along the *a*-axis penetrating through the $NiC_2$ atomic layer.

**Conclusion**

In summary, we investigated the crystal structure of $CeNiC_2$ from 0 to 18.6 GPa by using the single crystal X-ray diffraction with a laboratory X-ray source. Our results reveal a large Bulk modulus ~ 134 GPa and anisotropic linear compressibility following the relationship $|k_a| > |k_c| > |k_b|$. Although, we do not detect any signature of structural phase transition, the direction of FNN and SNN between the Ce-Ce and Ni-Ni atoms interchange, near the pressure where the antiferromagnetic ordering temperature reaches a maximum in the pressure temperature phase diagram of $CeNiC_2$. Our results suggest that the direction of nearest neighbors interexchange might play a key role in the suppression of magnetic order and the enhancement of Kondo effect.

**Acknowledgements**

We would like to thank Mingxuan Fu for useful comments and S. Nagasaki for help during experiments. This work was financially supported by JSPS KAKENHI Grant Number JP19H00648.